# THE GALAXY-HALO CONNECTION:

# PRESENT STATUS AND FUTURE PROSPECTS

> *Let us get away from Creation back to problems that we may possibly know something about.*
>
> Eddington (1933)

## 1. INTRODUCTION

To know something about the early Universe, one can either follow in the footsteps of Edwin Hubble and look back in time at the most distant galaxies, or one can follow in those of George Preston and search for the most metal-poor stars.

## 2. FORMATION OF THE HALO

During the period 1962-1977 thinking on Galactic evolution was guided by the seminal paper of Eggen, Lynden-Bell & Sandage (1962, ELS) in which it was suggested that the Milky Way System formed by the rapid dissipational collapse of a single protogalaxy. Faith in this paradigm was severely shaken by two papers presented at the 1977 Yale conference on *The Evolution of Galaxies and Stellar Populations*. At this meeting, Leonard Searle (1977) showed that, contrary to the expectation from the ELS model, globular clusters in the outer halo did *not* exhibit a radial abundance gradient. This conclusion is supported by modern data on RR Lyrae stars (Suntzeff, Kinman & Kraft 1991) and on globular clusters (van den Bergh 1995a). Furthermore, Alan Toomre (1977) remarked that "It seems almost inconceivable that there wasn't a great deal of merging of sizable bits and pieces (including quite a few lesser galaxies) early in the career of every major galaxy". These ideas of Searle and Toomre led Searle & Zinn (1978, SZ) to conclude that the outer halo of the Galaxy was formed from "transient protogalactic fragments that continued to fall into dynamical equilibrium with the Galaxy for some time after the collapse of its central regions had been completed". For more recent versions of this view, the reader is referred to van den Bergh (1993) and Zinn (1993). Observations which seem to indicate that the Galactic halo is not dynamically mixed (Majewski - this volume) appear consistent with the view that captured fragments contributed significantly to the formation of the halo. A possible problem with the Searle-Zinn scenario, in which the outer halo was formed by infall of "bits and pieces", is that it does not appear to account for the observation (van den Bergh 1995a) that the half-light radii of globular clusters in the outer Galactic halo increase with Galactocentric distance. The fact that the globular clusters associated with the Fornax dSph galaxy are smaller [ $<r_h> = 3.2$ pc] than most outer halo clusters (van den Bergh 1994) also seems to militate against the suggestion that a significant fraction of the outer Galactic halo was formed from the debris of Fornax-like dwarf galaxies.

An eloquent defence of the ELS paradigm has been given by Sandage (1989). Furthermore, the observation that the distant clusters NGC 2419 and Pal. 3 have ages close to those of the globulars near the Galactic center suggest that something rather coherent happened ~ 15 Gyr ago (Harris - this volume).

Observations of 11 Galactic globular clusters on retrograde orbits (van den Bergh 1993) yield a mean metallicity $<[Fe/H]> = 1.59 \pm 0.07$. This is significantly higher than the value $<[Fe/H]> = 1.86 \pm 0.08$ that is found for 13 LMC globular clusters listed by Suntzeff (1992). Since mean cluster metallicity increases with parent galaxy luminosity, this suggests that the Galactic globular clusters in retrograde orbits *might* have been formed in one or more ancestral galaxies with masses greater than that of the LMC, that were tidally captured into the Galactic halo from retrograde orbits.

The Searle-Zinn paradigm correctly predicts that the metallicity [Fe/H] of globular clusters in the outer halo should not correlate with Galactocentric distance $R_{GC}$. However, it is difficult to see how the SZ model can account for the observation that the half-light radii $r_h$ of outer halo clusters *do* correlate with $R_{GC}$. Specifically, van den Bergh (1995a) finds that the rank correlation coefficient between $r_h$ and $R_{GC}$, for clusters with $R_{GC} > 20$ kpc, is $\rho = + 0.61 \pm 0.18$. For these same clusters, an even stronger correlation $\rho = + 0.83 \pm 0.08$ is found between $r_h$ and the cluster perigalactic distance P. On the hypothesis that the halo represents material that fell into the Galaxy at a late date, it is also difficult to understand why the clusters NGC 2419 and Pal. 3, located at $R_{GC} \sim 100$ kpc, have ages (Richer et al. 1996) that are very similar to those of most clusters located at $R_{GC} < 10$ kpc.

## 3.  FORMATION OF THE BULGE

The Galaxy must have been formed with a hierarchy of collapse timescales t, in which

$$t \approx (G\rho)^{-0.5} . \qquad (1)$$

Since the density $\rho$ of the Galaxy increases towards its center, one would expect the core of the Milky Way System to be older than its outer regions. At first sight Baade's (1951) discovery of larger numbers of RR Lyrae variables in the direction of the Galactic nuclear bulge seemed to confirm the hypothesis that the bulge was very old. This conclusion also appeared to be strengthened by the work of Lee (1992) who found the horizontal branch stars in the bulge to be $1.3 \pm 0.3$ Gyr *older* than their counterparts in the Galactic halo. If the first stars formed in the nuclear bulge are old, then the very metal-poor nuclear globular cluster NGC 6287 (Stetson & West 1994) might turn out to be one of the oldest objects in the Galaxy. However, it is not yet clear whether the most metal deficient objects seen in the Galactic nuclear bulge are residents of, or visitors to, the central region of the Galaxy.

Although Baade's (1951) work had shown that the Galactic nucleus contained metal-poor stars, it was subsequently demonstrated by Morgan (1959) that the *dominant* population of the Galactic nuclear bulge consists of strong-lined metal-rich stars. Where did the gas that formed these metal-rich bulge stars come from? Carney et al. (1990), and Wyse & Gilmore (1992) have argued convincingly that such stars were probably formed from the low angular momentum component of left-over halo gas that sank towards the Galactic center. Spectroscopic observations of K giants in Baade's Window yield [Mg/Fe] $\approx + 0.3$. This Mg excess, which is similar to that observed in elliptical galaxies, suggests (van den Bergh 1995b) that the gas from which stars formed in the bulge was enriched in $\alpha$ elements on a short time-scale. Radial velocity observations (Minniti 1995a) show a strong dependence of kinematics on abundance, with metal-poor stars having a high velocity dispersion and metal-rich objects exhibiting prograde rotation and a smaller velocity dispersion. However, it is not yet possible to decide if the bulge and halo formed as distinct components, or whether there is a smooth transition between them. The observation that M33 has a well-developed halo containing globular clusters (Schommer et al. 1991) and RR Lyrae stars (Pritchet 1988), but appears to contain little or no nuclear bulge (Bothun 1991), suggests that the strengths of the bulge and halo population components are *not* closely correlated. Minniti (1995b) has argued that the metal-rich globular clusters within 3 kpc of the nucleus are physically associated with the Galactic bulge population, rather than with the thick disk or inner halo. Ortolani et al. (1995) find that the metal-rich bulge clusters NGC 6528 and NGC 6553 have ages that are, within a few Gyr, equal to those of typical halo globulars. However, such age estimates depend critically on the adopted absolute magnitude calibration of RR Lyrae stars. This calibration (Bolte - this volume) is rendered uncertain by the possibility that $M_V$ (RR), at a given value of [Fe/H], differs systematically between globular clusters and the Field.

Van den Bergh (1957) has shown that the nuclear region of the Galaxy presently contains far less gas than would have been ejected by evolving stars over a Hubble time. This suggests that a significant fraction of the stellar population in the bulge consists of second generation stars that were formed from gaseous material that had previously been processed through bulge stars.

It is presently not known whether mergers resulting from dynamical frictions have contributed to the stellar population in the nuclear bulge of the galaxy. Only mergers with massive companions could have contributed significantly to the strong population component with [Fe/H] > 0 that is observed in the nuclear bulge of the Galaxy.

## 4. FORMATION OF THE DISK

In his contribution to the Vatican *Conference on Stellar Populations*, Oort (1958) assigned stars to (1) Halo Population II, (2) Intermediate Population II, and (3) Disk Population I. In modern terminology, these components correspond (approximately) to (a) the Halo, (b) the Thick Disk, and (c) the Thin Disk. Excellent reviews of problems related to the structure, chemical composition, and evolution of these Galactic components are given by Gilmore, Wyse & Kuijken (1989), and in Majewski (1993). At the present time, it is still not entirely clear if the Halo, the Thick Disk and the oldest component of the Thin Disk form a continuum, or whether they are composed of stars formed during distinct Galactic evolutionary phases.

The disk of the Galaxy is a rapidly rotating dissipational structure that lost energy via cloud-cloud collisions. Hartwick (1976) pointed out that much of the gas in this disk probably originated in star forming regions of the halo, where some heavy element enrichment had already taken place. The Thin Disk represents the final phase of the dissipational settling of the halo gas. The evolutionary relationships between the Halo, Thick Disk and Thin Disk remain a subject of lively discussion. In particular, it is not yet clear whether evolution from the halo to disk phases of Galactic evolution was continuous, or if it was interrupted by the violent burst of star formation that must have occurred during the initial collapse of the Galactic halo (Berman & Suchkov 1991). The Large Magellanic Cloud is an example of a galaxy in which there seems to have been a particularly long hiatus between the initial burst of halo star formation (which produced 13 globular clusters and an old stellar population containing RR Lyrae stars) and a phase of disk star formation that started ~ 6 Gyr ago. Perhaps studies of Mira stars, which occur in great abundance in both the Halo and Thick Disk, could throw some light on the question whether there was a break between the Halo and Thick Disk phases of Galactic evolution.

Presently, only weak constraints can be placed on the age of the Galactic disk. NGC 6791, which is the one of the oldest known disk clusters, has an age of 7 - 10 Gyr (Kaluzny & Rucinski 1995, Tripicco et al. 1995). A number of possibly even older open clusters have been listed by Friel (1995). A trigonometric parallax for the white dwarf ESO 439-26 gives $M_V = +17.6 \pm 0.1$, which yields a cooling age of 11.5 Gyr for a C-core white dwarf, or a cooling age of 8.5 Gyr for an O-core white dwarf of 0.6 $M_\odot$ (Wood 1995). These values are, of course, *lower limits* to the age of the Galactic disk. Upper limits to the age of the disk are set by the ~ 16 Gyr age of the oldest globular clusters (Bolte - this volume) and by the 16 ± 5 Gyr age of the supernova that produced the thorium that is observed (Sneden et al. - this volume) in the extremely metal-poor star CS22892-052.

Majewski (1993) gives a table listing key features of eight scenarios that have been suggested for the formation of the Thick Disk. Among these one might mention: (1) the Thick Disk forms when pressure support starts at the beginning of the dissipative phase, (2) the Thick Disk is formed by violent heating of the Thin Disk by satellite accretion and (3) the Thick Disk is assembled by direct accretion of disrupted satellite galaxies. The observation (Gratton & Carretta - this volume) that [Mg/Fe] ~ 0.4 in the thick disk, but that it decreases by 0.2 dex at the thick-thin disk transition, constrains scenarios that can be invoked to explain the formation of the thick disk. Possibly, systematics of nearby edge-on spiral galaxies that do, or do not, have thick disks will eventually throw some light on this intriguing problem. The very limited data that are available so far (Morrison - this volume) suggest that only galaxies with nuclear bulges exhibit thick disks. Important constraints on scenarios for the formation of the Thick Disk would be provided by more data on the metallicity range that is spanned by stars belonging to the both the Thick Disk and Thin Disk. It would also be very helpful to have more information on the radial and vertical abundance gradients, and on the radial extent of, the Thick Disk. The existence of a very metal-poor Thick Disk component now appears to be in doubt (Twarog & Anthony-Twarog 1996). Realistic capture computations (Barnes - this volume) suggest that it may not be possible to account for very metal-poor disk stars by invoking mergers with metal-poor satellites.

## 5.     CONCLUSIONS

Over the last decade, a picture of Galactic evolution has emerged in which the central region of the protogalaxy collapsed on a relatively short time-scale *à la* ELS, whereas the outer part of the Halo was assembled over a longer period of time, as originally envisaged by SZ. Most of the data presented at this meeting appear to fit comfortably within such a paradigm. However, there are still a few worries. In my view, the most serious is that our present "Standard Model" does not appear to account for two observations: (1) the clusters NGC 2419 and Pal. 3 at $R_{GC} \approx 100$ pc have ages (Richer et al. 1996) that are only ~ 1 Gyr younger than those of the oldest clusters at small values of $R_{GC}$. (2) It is not obvious how the SZ picture can account for the observation (van den Bergh 1994) that the half-light diameters of globular clusters in the outer halo increase with $R_{GC}$. Nor does the "Standard Model" account for the complete absence of compact (and hence dynamically very stable) globular clusters in the outer halo of the Galaxy. Finally, (3) it is not clear why globular cluster ages and cluster diameters (van den Bergh 1995b) correlate much better with their perigalactic distances than they do with their present Galactocentric distances $R_{GC}$. Hopefully, new observations will throw more light on these and other problems. However, it might be that, in the words of Erwin Schrödinger, "*the task is, not so much to see what no one has yet seen; but to think what nobody has yet thought, about that which everybody sees*."